# Compendium of vector analysis
# with applications to continuum mechanics

compiled by **Valery P. Dmitriyev**

*Lomonosov University*
*P.O.Box 160, Moscow 117574, Russia*
*e-mail: dmitr@cc.nifhi.ac.ru*

## 1. Connection between integration and differentiation

### Gauss-Ostrogradsky theorem

We transform the volume integral into a surface one:

$$\int_V \partial_i P dV = \int_V \partial_i P dx_i dx_j dx_k = \int_{S(V)} dx_j dx_k \Big|_{x_i^-(x_j,x_k)}^{x_i^+(x_j,x_k)} P =$$

$$= \int_{S(V)} dx_j dx_k \left[ P(x_i^+(x_j,x_k), x_j, x_k) - P(x_i^-(x_j,x_k), x_j, x_k) \right] =$$

$$= \int_{S^+} \cos\theta_{ext}^+ dSP - \int_{S^-} \cos\theta_{int}^- dSP = \oint_S \cos\theta_{ext} dSP = \oint \mathbf{n}\cdot\mathbf{e}_i P dS$$

Here the following denotations and relations were used:
$P$ is a multivariate function $P(x_i, x_j, x_k)$, $\partial_i = \partial/\partial x_i$, $V$ volume,
$S$ surface, $\mathbf{e}_i$ a basis vector, $\mathbf{e}_i \cdot \mathbf{e}_j = \delta_{ij}$, $\mathbf{n}$ the external normal to the element $dS$ of closed surface with

$$dx_j dx_k = |\mathbf{n}\cdot\mathbf{e}_i| dS, \quad \mathbf{n}\cdot\mathbf{e}_i = \cos\theta .$$

Thus

$$\int_V \partial_i P dV = \oint_{S(V)} P \mathbf{n}\cdot\mathbf{e}_i dS \qquad (1.1)$$

Using formula (1.1), the definitions below can be transformed into coordinate representation.

**Gradient**
$$\oint_{S(V)} P\mathbf{n}\,dS = \oint_{S(V)} (\mathbf{n}\cdot\mathbf{e}_i)\mathbf{e}_i P\,dS = \int_V \partial_i P\mathbf{e}_i\,dV$$

where summation over recurrent index is implied throughout. By definition
$$\operatorname{grad} P = \nabla P = \partial_i P\mathbf{e}_i$$

**Divergence**
$$\oint_{S(V)} \mathbf{A}\cdot\mathbf{n}\,dS = \oint_{S(V)} (\mathbf{n}\cdot\mathbf{e}_i) A_i\,dS = \int_V \partial_i A_i\,dV \quad (1.2)$$

By definition
$$\operatorname{div}\mathbf{A} = \nabla\cdot\mathbf{A} = \partial_i A_i$$

**Curl**
$$\oint_{S(V)} \mathbf{n}\times\mathbf{A}\,dS = \oint_{S(V)} (\mathbf{n}\cdot\mathbf{e}_i)\mathbf{e}_i\times A_j\mathbf{e}_j\,dS = \int_V \partial_i A_j\mathbf{e}_i\times\mathbf{e}_j\,dV \quad (1.3)$$

By definition
$$\operatorname{curl}\mathbf{A} = \nabla\times\mathbf{A} = \partial_i A_j \mathbf{e}_i\times\mathbf{e}_j$$

**Stokes theorem** follows from (1.3) if we take for the volume a right cylinder with the height $h\to 0$. Then the surface integrals over the top and bottom areas mutually compensate each other. Next we consider the triad of orthogonal unit vectors
$$\mathbf{m},\ \mathbf{n},\ \boldsymbol{\tau}$$

where $\mathbf{m}$ is the normal to the top base and $\mathbf{n}$ the normal to the lateral face
$$\boldsymbol{\tau}=\mathbf{m}\times\mathbf{n}$$

Multiplying the left-hand side of (1.3) by $\mathbf{m}$ gives
$$\int_{lateral}\mathbf{m}\cdot(\mathbf{n}\times\mathbf{A})\,dS = \int_{lateral}(\mathbf{m}\times\mathbf{n})\cdot\mathbf{A}\,dS = \int_{lateral}\boldsymbol{\tau}\cdot\mathbf{A}\,dS = h\oint_l \boldsymbol{\tau}\cdot\mathbf{A}\,dl$$

where $\boldsymbol{\tau}$ is the tangent to the line. Multiplying the right-hand side of (1.3) by $\mathbf{m}$ gives
$$h\int_S \mathbf{m}\cdot\operatorname{curl}\mathbf{A}\,dS$$

where $\mathbf{m}$ is the normal to the surface. Now, equating both sides, we come to the formula sought for
$$\oint_l \boldsymbol{\tau}\cdot\mathbf{A}\,dl = \int_S \mathbf{m}\cdot\operatorname{curl}\mathbf{A}\,dS$$

The Stokes theorem is easily generalized to a nonplanar surface (applying to it Ampere's theorem). In this event, the surface is approximated by a polytope. Then mutual compensation of the line integrals on common borders is used.

# 2. Elements of continuum mechanics

A medium is characterized by the volume density $\rho(\mathbf{x},t)$ and the flow velocity $\mathbf{u}(\mathbf{x},t)$.

**Continuity equation**

The mass balance in a closed volume is given by

$$\partial_t \int_V \rho dV + \oint_{S(V)} \rho \mathbf{u}\cdot\mathbf{n} dS = 0$$

where $\partial_t = \partial/\partial t$. We get from (1.2)

$$\oint \rho \mathbf{u}\cdot\mathbf{n} dS = \int \partial_i(\rho u_i) dV$$

Thereof the continuity equations follows

$$\partial_t \rho + \partial_i(\rho u_i) = 0$$

**Stress tensor**

We consider the force $d\mathbf{f}$ on the element $dS$ of surface in the medium and are interested in its dependence on normal $\mathbf{n}$ to the surface

$$d\mathbf{f}(\mathbf{n})$$

where

$$d\mathbf{f}(-\mathbf{n}) = -d\mathbf{f}(\mathbf{n})$$

With this purpose the total force on a closed surface is calculated. We have for the force equilibrium at the coordinate tetrahedron

$$d\mathbf{f}(\mathbf{n}) + d\mathbf{f}(\mathbf{n}_1) + d\mathbf{f}(\mathbf{n}_2) + d\mathbf{f}(\mathbf{n}_3) = 0$$

where the normals are taken to be external to the surface

$$\mathbf{n}_1 = -sign(\mathbf{n}\cdot\mathbf{e}_1)\mathbf{e}_1, \quad \mathbf{n}_2 = -sign(\mathbf{n}\cdot\mathbf{e}_2)\mathbf{e}_2, \quad \mathbf{n}_3 = -sign(\mathbf{n}\cdot\mathbf{e}_3)\mathbf{e}_3$$

Thence

$$d\mathbf{f}(\mathbf{n}) = sign(\mathbf{n}\cdot\mathbf{e}_j) d\mathbf{f}(\mathbf{e}_j) \qquad (2.1)$$

The force density $\boldsymbol{\sigma}(\mathbf{n})$ is defined by

$$d\mathbf{f} = \boldsymbol{\sigma} dS$$

Insofar as

$$dS_j = |\mathbf{n}\cdot\mathbf{e}_j| dS$$

we have for (2.1)

$$d\mathbf{f}(\mathbf{n}) = sign(\mathbf{n}\cdot\mathbf{e}_j)\boldsymbol{\sigma}(\mathbf{e}_j)dS_j = sign(\mathbf{n}\cdot\mathbf{e}_j)|\mathbf{n}\cdot\mathbf{e}_j|\boldsymbol{\sigma}(\mathbf{e}_j)dS = \mathbf{n}\cdot\mathbf{e}_j\boldsymbol{\sigma}(\mathbf{e}_j)dS$$

i.e.

$$\boldsymbol{\sigma}(\mathbf{n}) = \mathbf{n}\cdot\mathbf{e}_j\boldsymbol{\sigma}(\mathbf{e}_j)$$

$$= \mathbf{n}\cdot\mathbf{e}_j\mathbf{e}_i\sigma_i(\mathbf{e}_j)$$

The latter means that $\boldsymbol{\sigma}(\mathbf{n})$ possesses the tensor property. The elements of the stress tensor are defined by

$$\sigma_{ij} = \sigma_i(\mathbf{e}_j)$$

Now, using (1.2), the force on a closed surface can be computed as a volume integral

$$\oint \boldsymbol{\sigma}(\mathbf{n})dS = \oint \boldsymbol{\sigma}(\mathbf{e}_j)\mathbf{e}_j\cdot\mathbf{n}dS = \int_V \partial_j\boldsymbol{\sigma}(\mathbf{e}_j)dV \qquad (2.2)$$

**Euler equation**

The momentum balance is given by the relation

$$\partial_t \int_V \rho\mathbf{u}dV + \oint_{S(V)}(\rho\mathbf{u})\mathbf{u}\cdot\mathbf{n}dS = \oint_{S(V)}\boldsymbol{\sigma}dS \qquad (2.3)$$

We have for the second term by (1.2)

$$\oint (\rho\mathbf{u})\mathbf{u}\cdot\mathbf{n}dS = \oint (\rho\mathbf{u})u_j\mathbf{e}_j\cdot\mathbf{n}dS = \int \partial_j(\rho u_j\mathbf{u})dV$$

Using also (2.2) gives for (2.3)

$$\partial_t(\rho\mathbf{u}) + \partial_j(\rho u_j\mathbf{u}) = \partial_j\boldsymbol{\sigma}(\mathbf{e}_j)$$

or

$$\rho\partial_t\mathbf{u} + \rho u_j\partial_j\mathbf{u} = \partial_j\boldsymbol{\sigma}(\mathbf{e}_j) \qquad (2.4)$$

**Hydrodynamics**

The stress tensor in a fluid is defined from the pressure as
$$\sigma_{ij} = -p\delta_{ij}$$
That gives for (2.4)
$$\rho\partial_t u_i + \rho u_j \partial_j u_i + \partial_j p = 0$$

**Elasticity**

The solid-like medium is characterized by the displacement $\mathbf{s}(\mathbf{x},t)$. For small displacements
$$\mathbf{u} = \partial_t \mathbf{s}$$
and the quadratic terms in the left-hand part of (2.4) can be dropped. For an isotropic homogeneous medium the stress tensor is determined from the Hooke's law as
$$\sigma_i(\mathbf{e}_j) = \lambda \delta_{ij} \partial_k s_k + \mu(\partial_i s_j + \partial_j s_i)$$
where $\lambda$ and $\mu$ are the elastic constants. That gives
$$\partial_j \sigma_i(\mathbf{e}_j) = \lambda \partial_i \partial_k s_k + \mu(\partial_i \partial_j s_j + \partial_j^2 s_i) = (\lambda+\mu)\partial_i \partial_j s_j + \mu \partial_j^2 s_i$$
and
$$\partial_j \boldsymbol{\sigma}(\mathbf{e}_j) = (\lambda+\mu)\operatorname{graddiv}\mathbf{s} + \mu\nabla^2 \mathbf{s}$$
$$= (\lambda+2\mu)\nabla^2 \mathbf{s} + (\lambda+\mu)\operatorname{curlcurl}\mathbf{s}$$
$$= \lambda\operatorname{graddiv}\mathbf{s} - \mu\operatorname{curlcurl}\mathbf{s}$$
where $\operatorname{graddiv} = \nabla^2 + \operatorname{curlcurl}$ was used. Substituting it to (2.4) we get finally Lame equation
$$\rho\partial_t^2 \mathbf{s} = (\lambda+\mu)\operatorname{graddiv}\mathbf{s} + \mu\nabla^2 \mathbf{s}$$
where $\rho$ is constant.